\documentclass[aps,prb,showpacs, twocolumn,floatfix,superscriptaddress]{revtex4-1}
\usepackage{graphicx,hyperref,bbm,amssymb,amsfonts,amsmath,times,color,soul,epstopdf}  

\date{\today}

\begin{document}

\title{Dimensional crossover and the link between thermodynamics and dynamics:\ the case of Ising models at complex temperature}

\author{Sankhya Basu}
\affiliation{Physics program and Initiative for the Theoretical Sciences, The Graduate Center, CUNY, New York, NY 10016, USA}
\affiliation{Department of Physics and Astronomy, College of Staten Island, CUNY, Staten Island, NY 10314, USA}
\author{Chris A. Hooley}
\affiliation{SUPA, School of Physics and Astronomy, University of St Andrews, North Haugh, Fife KY16 9SS, United Kingdom}
\author{Vadim Oganesyan}
\affiliation{Physics program and Initiative for the Theoretical Sciences, The Graduate Center, CUNY, New York, NY 10016, USA}
\affiliation{Department of Physics and Astronomy, College of Staten Island, CUNY, Staten Island, NY 10314, USA}

\date{Tuesday 21st April 2020}
\begin{abstract}
We study dimensional crossover in Ising systems at complex temperatures by comparing three types of system:\ the infinite isotropic 2D Ising model; the infinite anisotropic 2D Ising model; and Ising ladders with a finite number of legs.  In particular we present evidence, from both tensor-network calculations and numerical evaluations based on the exact solution of the model, that the infinite anisotropic 2D Ising model exhibits long-range spatially modulated magnetization in certain regions of the complex-temperature plane.  We discuss the physics of the special unitary points that exists in the complex-temperature plane, and their connections to the theory of quantum information processing.
\end{abstract}

\maketitle

{\it Introduction.}
The importance of dimensionality for the equilibrium properties of many-particle systems, especially the stability or otherwise of long-range-ordered phases, has been known now for several decades.  The Mermin-Wagner theorem\cite{MW} established that the spontaneous breaking of a continuous symmetry can in general not occur at all in $d=1$, and can occur only at $T=0$ in $d=2$.  The `Landau argument'\cite{Landau} further demonstrated that the spontaneous breaking of a discrete symmetry can occur only at $T=0$ in $d=1$, while in $d \geqslant 2$ finite-temperature transitions are permitted.

Berezinskii\cite{Berezinskii} and Kosterlitz and Thouless\cite{KT} showed that systems with O(2) symmetry in $d=2$ constitute an important special case:\ while, as per the Mermin-Wagner theorem, they cannot exhibit spontaneous symmetry breaking at $T>0$, they can however show `topological order' --- which results in power-law decay of correlations between the spins --- in a temperature range $0 < T < T_{\rm BKT}$.

It is thus natural to ask how long-range order emerges as the system in question interpolates between (for example) $d=1$ and $d=2$, and there is a considerable literature addressing this question for various classes of system.  Such interpolation can be achieved in many ways, including coupling $d=1$ chains into finite-width ladders\cite{ladders}, or continuously tuning the ratio between the horizontal and vertical couplings in a $d=2$ lattice system from 0 to 1\cite{anisotropic2d}.  While the end-points of these approaches match by construction, the physics in between these end-points is both subtle and dependent on the particular interpolation scheme used.

\begin{figure}
\begin{center}
\includegraphics[width=0.9\columnwidth]{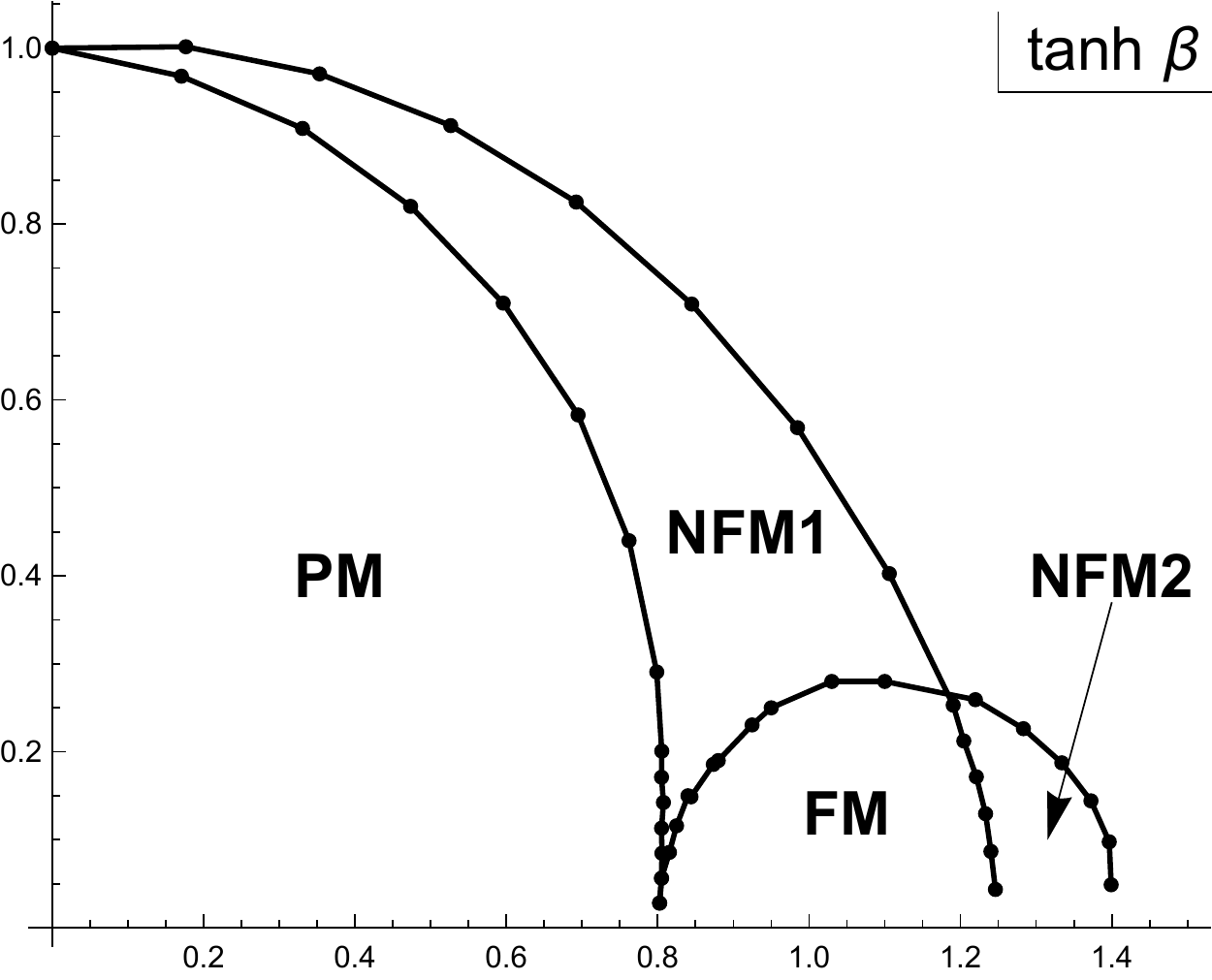}
\end{center}
\caption{Phase diagram of the anisotropic ($J_y/J_x=1/10$) 2D Ising model at complex temperature in the thermodynamic limit.  The points represent the locations of phase transitions as determined from numerical calculations of the type presented in Fig.~\ref{f:energyspecheat}; the lines are guides to the eye.  As well as the ferromagnetic `bubble' around the zero-temperature point and the disordered regions near the origin and at large $\vert {\tanh \beta} \vert$, there are two additional non-ferromagnetic phases:\ `NFM1' and `NFM2'.  We believe that these phases exhibit long-range spatially modulated spin-spin correlations.}
\label{f:phasediag}
\end{figure}

Over the past few decades, considerable light has been shed on several open problems in statistical mechanics by the complex temperature technique, pioneered by Michael Fisher and others\cite{complextemp}.  In this approach, one considers the inverse temperature $\beta$ in the partition function $Z(\beta) = \sum_C \exp \left( - \beta E_C \right)$ formally to be a complex variable.  This opens up the possibility of points in the complex $\beta$-plane where the finite-size lattice partition function vanishes, something which is not possible for real temperature.  For simple models, such as the isotropic 2D Ising model in the thermodynamic limit, these `Fisher zeros' form contours in the complex $\beta$-plane, which cut the real $\beta$-axis at the temperature corresponding to the phase transition.  If the density of zeros vanishes as the real $\beta$-axis is approached, the transition is continuous; if the density remains finite, the transition is first-order.

The complex temperature technique, and related techniques such as the complex magnetic field technique due to Lee and Yang\cite{leeyang, yanglee}, can do more than just identify the location and order of phase transitions.  They provide an elegant method for deriving the critical exponents\cite{deger} and the scaling relations between them that hold near continuous phase transitions, as well as allowing the derivation of analogous and previously unknown relations for the logarithmic corrections to scaling\cite{kenna}.  Despite these successes, however, the complex temperature technique has mainly been seen as an essentially formal one.

In this paper, we explore an alternative view:\ that the contours of Fisher zeros should be viewed as {\it bona fide\/} phase transitions, and the regions between them as {\it bona fide\/} phases.  In particular, we present evidence that some of the models that interpolate between the $d=1$ Ising chain and the $d=2$ square-lattice Ising model have modulated magnetic phases, which we refer to as `non-ferromagnets' (NFM), in their complex-temperature phase diagrams, in addition to the ferromagnetic and short-range-ordered phases that would be expected from a real-temperature analysis.  An example of such a phase diagram is shown in Fig.~\ref{f:phasediag}.

Furthermore, we show that for all of the models we consider there are special `unitary points' in the complex-temperature phase diagram.  At these points, the transfer matrix of the statistical mechanical model becomes unitary, and thus may be interpreted directly as representing the dynamics of an isolated quantum system.  This complements other intriguing suggestions in the recent literature of links between quantum dynamics and complex-temperature thermodynamics\cite{heyl}.

{\it Notation and methods.}
We adopt units in which $J_x = 1$, where $J_x$ is either the coupling constant (in the case of isotropic models) or the coupling constant in the more strongly coupled direction (in the case of anisotropic models).  Furthermore, we often use the phrase `complex temperature' as shorthand for the complex plane of hyperbolic tangent of inverse temperature, $\tanh \beta$.  This variable is natural both in one-dimensional\cite{Beichert2013} and two-dimensional\cite{Fisher1965} problems.

We examine only the real part of the dimensionless free energy density ${\rm Re}(\beta f) \equiv - {\rm Re}(\ln Z)/A$, where $A$ is the total number of spins\footnote{It is an interesting question how to consistently define the rest of the complex free energy}.  From this we define a set of proxies to the analytic continuations of familiar thermodynamic functions:\ the free energy density $f \equiv {\rm Re}(\beta f)/\beta$, the internal energy density $u\equiv |\partial {\rm Re}(\beta f)/\partial \beta|$, and the specific heat capacity $C=|\partial u/\partial \beta|$.  Here $\partial/\partial \beta$ denotes a derivative taken along a radial line in the $\tanh \beta$ plane.

We employ a combination of exact and approximate methods to obtain our results:\ numerically exact transfer matrix diagonalization to obtain the free energy and information about the correlations in ladders in an arbitrary uniform magnetic field; analytic continuation of the result of Onsager's (partially\footnote{The exact solutions of the Ising model by Onsager and several others in the years that followed\cite{anisotropic2d} usually consist of several contributions only one of which dominates in the thermodynamic limit.  Fisher's original argument for generalizing Yang-Lee results was based on a seemingly incorrect procedure whereby he analytically continued only the part of the result that was important at real temperature. As explicitly demonstrated by Beichert {\it et al.\/}, this produces an entirely wrong pattern of zeros in ladders. Remarkably, Fisher's approximate solution is accurately reproduced by the unbiased TRG\cite{LevinNave} computational scheme.}) exact solution of the 2D Ising model; and Tensor Renormalization Group (TRG)\cite{LevinNave,GarciaWei} calculations of the real part of the free energy density, again in an arbitrary uniform magnetic field.

\begin{figure}
\begin{center}
\includegraphics[width=0.95\columnwidth]{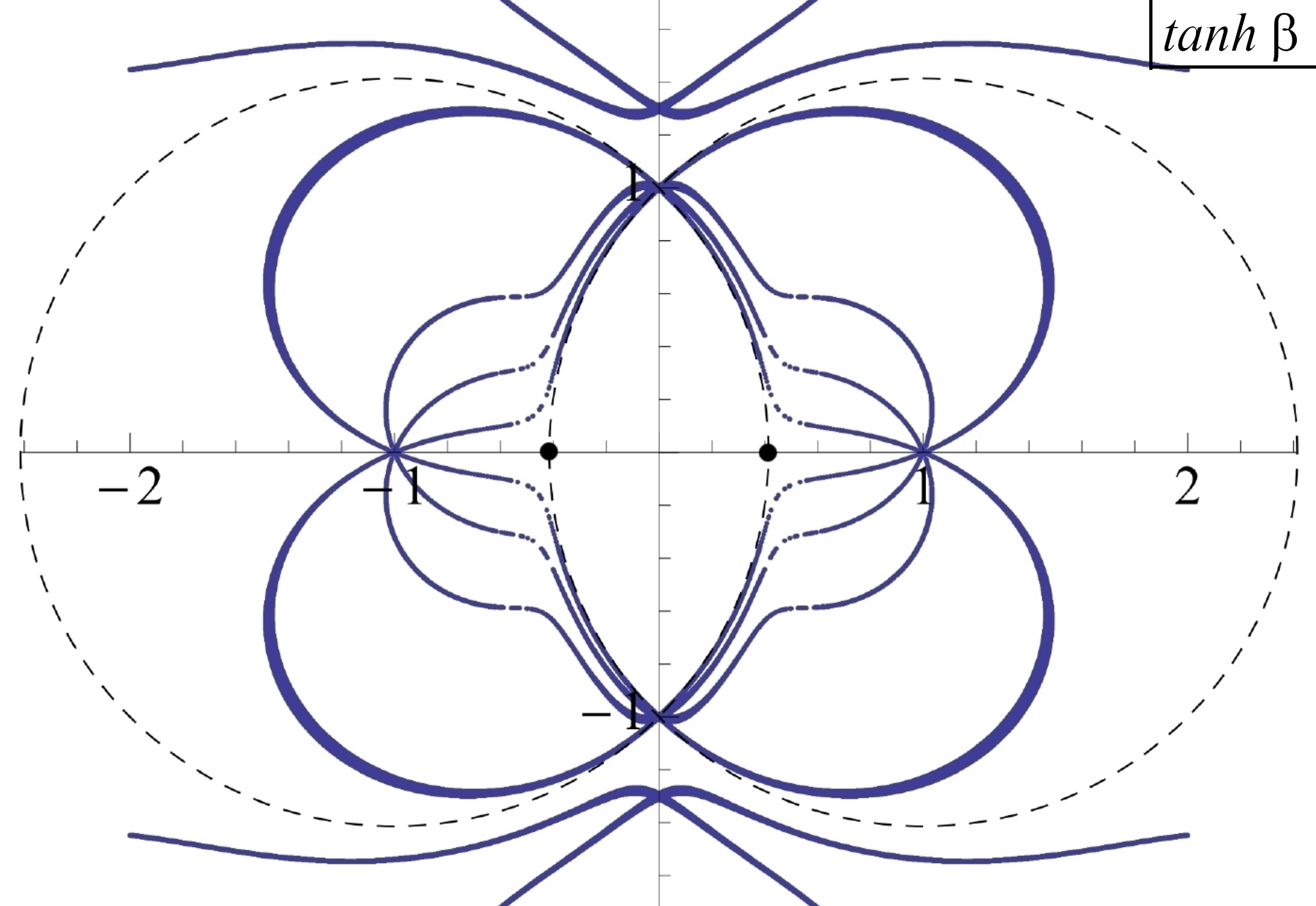}
\end{center}
\caption{(Reproduced from F.~Beichert {\it et al.}\cite{Beichert2013}.) The contours containing the zeros of the partition function in the complex temperature plane for Ising spin ladders with 4 legs and isotropic spin-spin interactions ($J_x = J_y$).  Dashed lines represent the expected location of Fisher zeros of the isotropic 2D Ising model\cite{Fisher1965}. The total number of contours is equal to the number of legs; their location was found to be only weakly sensitive to boundary conditions in the short (rung) direction and insensitive to the boundary conditions in the long (infinite) direction.}
\label{f:laddersfeli}
\end{figure}

{\it Ising ladders.}
We begin with the 1D Ising chain, which is the 1D limit of both the finite-width ladder and the anisotropic 2D Ising model.  As pointed out by Beichert {\it et al.\/}\cite{Beichert2013}, the $N$-site Ising chain has $N$ partition function zeros on the unit circle in the complex $\tanh \beta$ plane; however, these are interleaved with $N$ points at which the spin-spin correlations in the chain exhibit long-range magnetic order with a finite wavenumber $q = {\rm arg} \left( \tanh \beta \right)$.  This sequence includes a ferromagnetic ($q=0$) point at zero temperature for positive $J$ (i.e.\ $\tanh \beta = 1$) and a N{\'e}el-ordered point ($q=\pi$) at zero temperature for negative $J$ (i.e.\ $\tanh \beta = -1$), but also has another $N-2$ points that cover all other wavenumbers consistent with the periodic boundary conditions.

What happens to (a) the partition function zeros, and (b) the long-range-ordered points as we move towards the isotropic 2D model?  One way to investigate this is to couple $M$ Ising chains to form an $M$-leg ladder, with $N$ spins per leg.  As shown in Fig.~\ref{f:laddersfeli}, this splits the original contour of zeros into $M$ contours\cite{Beichert2013}, which all converge at the two zero-temperature points ($\tanh \beta = \pm 1$), but also at two points on the imaginary-temperature axis ($\tanh \beta = \pm i$).  We discuss these `unitary points' below.

For any 1D Ising system, the only place where true long-range order can occur is on the same contours where the partition function zeros are located.  This is because the condition for the two is the same, viz.\ that the largest two eigenvalues of the transfer matrix be equimodular.  For finite $M$ these contours are a set of measure zero; furthermore, changing the boundary conditions along the long direction of the ladder can perturb or even altogether remove these Fisher zeros.  It is therefore important to develop an understanding of this problem that transcends the analysis of partition function zeros to concentrate on more robust notions, e.g.\ correlation lengths.

\begin{figure}
\begin{center}
$\qquad$ $M=2$, longest $\qquad \qquad$
$M=2$, second-longest
\includegraphics[width=0.9\columnwidth]{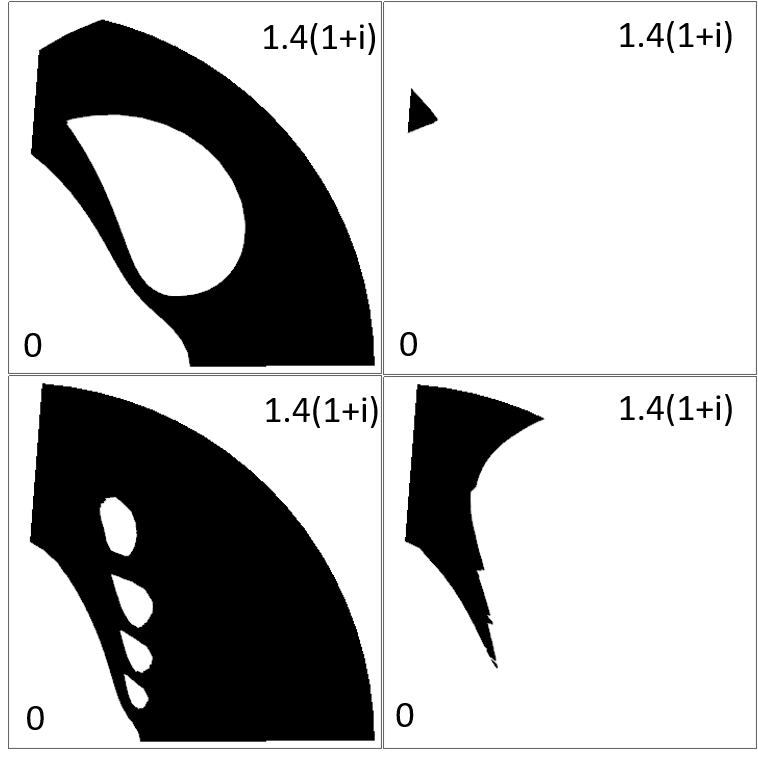}

\vspace{-1mm}
$\qquad$ $M=5$, longest $\qquad \qquad$
$M=5$, second-longest
\end{center}
\caption{The top-right quadrant of the complex $\tanh \beta$ plane of Fig.~\ref{f:laddersfeli} for an $M$-leg Ising spin ladder with isotropic interactions, $J_x = J_y = 1$.  Upper panels:\ for a 2-leg ladder, the longest correlation length (left) and the second-longest (right), both in the `long' direction, i.e.\ along the legs of the ladder.  Lower panels:\ the same for a 5-leg ladder.  The regions in which the correlation length in question exceeds 10 lattice sites are shaded black.  Note that, except near the unitary point $\tanh \beta = i$, only the longest correlation length shows significant structure.}
\label{f:ladders}
\end{figure}

With that in mind, we investigate what happens to the longest spin-spin correlation length (which is determined by the ratio between the largest and second-largest transfer-matrix eigenvalues) in the regions between the $M$ contours that `coalesce' into the ordered region of the 2D Ising model as $M \to \infty$.  We find that the correlation length stays very high, and grows as $M$ is increased:\ see the left-hand panels of Fig.~\ref{f:ladders}.
This suggests that, in the $M \to \infty$ limit, the two leading transfer-matrix eigenvalues become equimodular in a finite-area region.  Furthermore, in the thermodynamic limit their phases lock, resulting in a finite area of the complex-temperature plane in which the model exhibits ferromagnetic long-range order.  This region is just the area between the two `Fisher circles' shown by dashed lines in Fig.~\ref{f:laddersfeli}.  The second-longest correlation length (right-hand panels of Fig.~\ref{f:ladders}), by contrast, is short everywhere in this quadrant of the complex-temperature plane except near the unitary point, $\tanh\beta = i$.

\begin{figure}
\begin{center}
$\qquad$ $M=2$, longest $\qquad \qquad$
$M=2$, second-longest
\includegraphics[width=0.9\columnwidth]{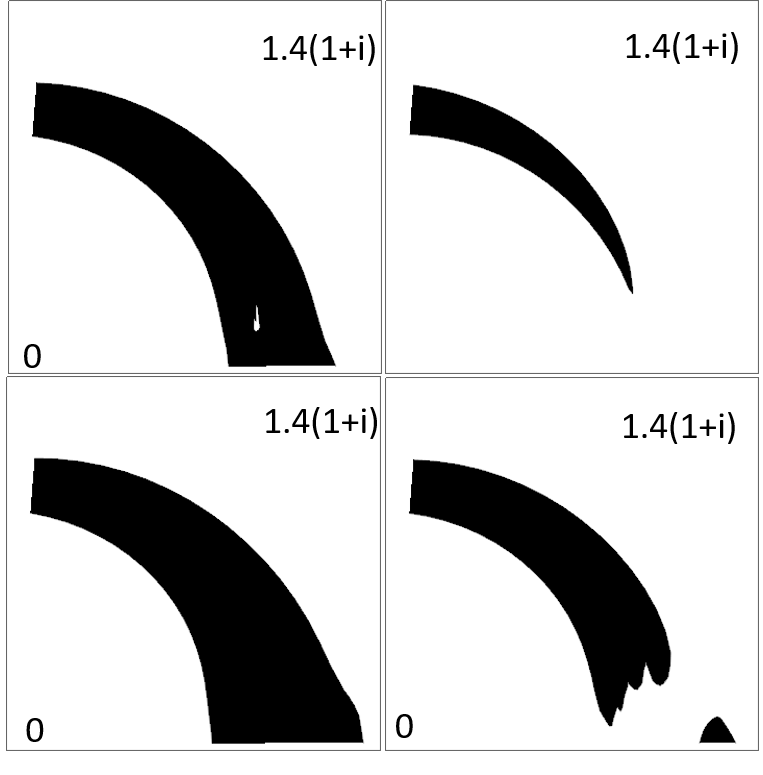}

\vspace{-1.8mm}
$\qquad$ $M=5$, longest $\qquad \qquad$
$M=5$, second-longest
\end{center}
\caption{The same as Fig.~\ref{f:ladders}, but for the case of anisotropic interactions, $J_y = J_x/10 = 1/10$.  Note that, in a region roughly corresponding to the `NFM1' and `NFM2' regions of Fig.~\ref{f:phasediag}, the second-longest correlation length also shows non-trivial structure.}
\label{f:laddersaniso}
\end{figure}
Thus it seems that the isotropic 2D Ising model at complex temperature has only one type of ordered phase:\ long-range ferromagnetism.  What happened to the modulated magnetic order that we saw in the 1D chain?  To answer this question, we broaden our studies to Ising ladders in which the couplings are anisotropic, with $J_y < J_x \equiv 1$.

In Fig.~\ref{f:laddersaniso}, we show correlation length plots similar to those of Fig.~\ref{f:ladders}, but now for the anisotropic case $J_y = 1/10$.  The left-hand panels clearly demonstrate that the tendency to long-range order survives over a wide region of the complex-temperature plane.  However, as the right-hand panels show, the sub-dominant correlation length also becomes large in the anisotropic case, in a region that corresponds roughly to the regions denoted `NFM1' and `NFM2' in Fig.~\ref{f:phasediag}.  We shall now present evidence that, in the $M \to \infty$ limit, these regions show long-range modulated magnetic order.


{\it Evidence for modulated magnetic phases.}
The complex-temperature phase transitions of the anisotropic 2D Ising model in the thermodynamic limit can be located by studying the specific heat capacity, the complex-temperature version of which is defined above.  The nature of the ordered phases can be probed by measuring the spontaneous uniform magnetization, using the usual definition $m_0 \equiv \lim_{h \to 0^{+}} \left[ m(h) \right]$.  We shall show below that the `NFM1' and `NFM2' regions (a) are separated from the high-temperature paramagnetic region by phase transitions, but (b) have zero spontaneous uniform magnetization.  On the basis of these facts, we argue that they are long-range ordered phases in which the magnetization is spatially modulated.

\begin{figure}
\begin{center}
\includegraphics[width=0.95\columnwidth]{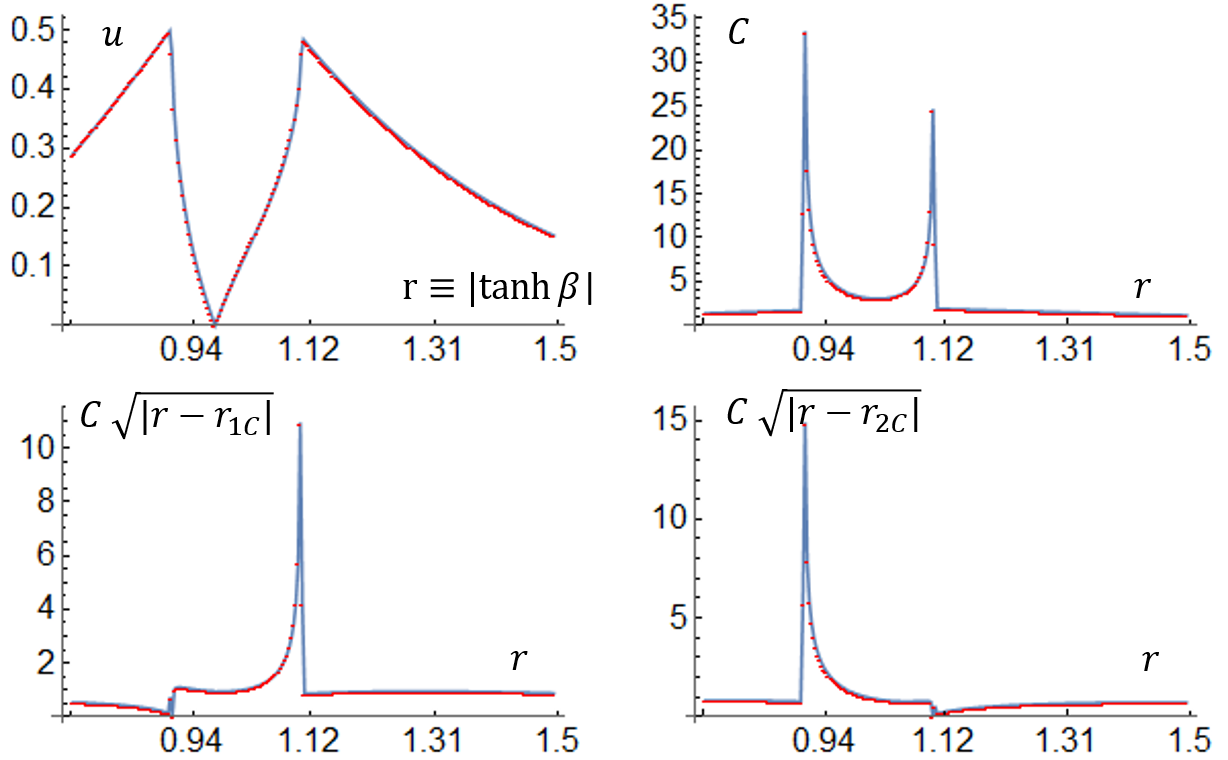}
\end{center}
\caption{Top left: The internal energy as a function of distance along a radial line at angle $2 \pi/9$ to the positive real axis in the complex $\tanh \beta$ plane.  The two transitions are those into and out of the `NFM1' phase (see Fig.~\ref{f:phasediag}).  Top right: The specific heat capacity along the same contour (see text for the precise definition of `specific heat capacity' at complex temperature).  Bottom panels: The same specific heat graph, but multiplied by the specified factor, demonstrating that the two singularities are one-sided square-root singularities.}
\label{f:energyspecheat}
\end{figure}

We begin by mapping out the phase boundaries of the anisotropic 2D Ising model in the complex-temperature plane using the numerical evaluation of the Onsager solution.  We have checked the results of this method against our tensor network computations of the free energy, and found good agreement.  Fig.~\ref{f:energyspecheat} shows our results for the internal energy (top left panel) and specific heat capacity (top right panel), along a radial line making an angle $\theta = 2\pi/9$ to the positive real axis in the complex $\tanh \beta$ plane.

The two phase transitions, marked by peaks in the specific heat capacity, contrast sharply with the isotropic case.  There are additional phase boundaries induced by anisotropy; further, these new phase transitions are not first-order, as those in the isotropic model are\cite{Fisher1965}.  Rather, they are characterized by one-sided square-root singularities in the specific heat, reminiscent of the commensurate-incommensurate transition\cite{Bak82}. Interestingly, these singularities appear at both the PM-NFM1 and FM-NFM1 transition boundaries, with the critical fluctuations and specific-heat divergence in the NFM1 phase.

\begin{figure}
\begin{center}
\includegraphics[width=0.9\columnwidth]{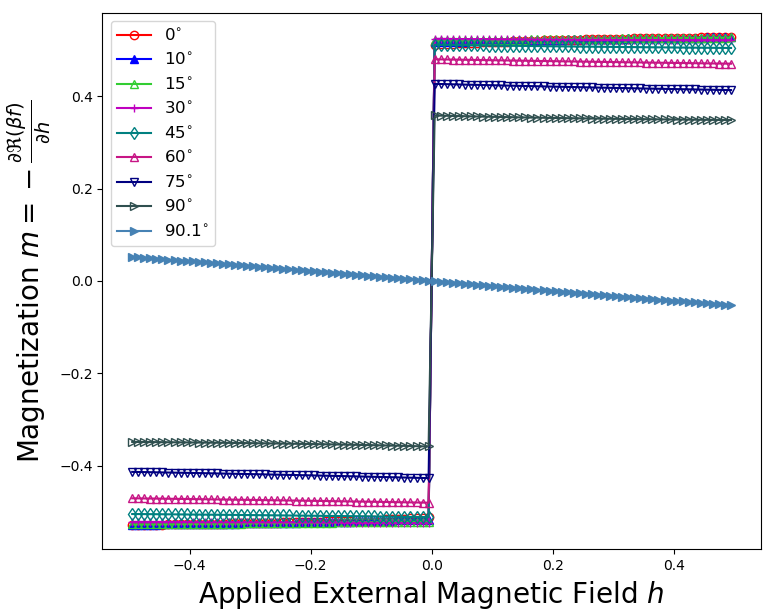}
\includegraphics[width=0.9\columnwidth]
{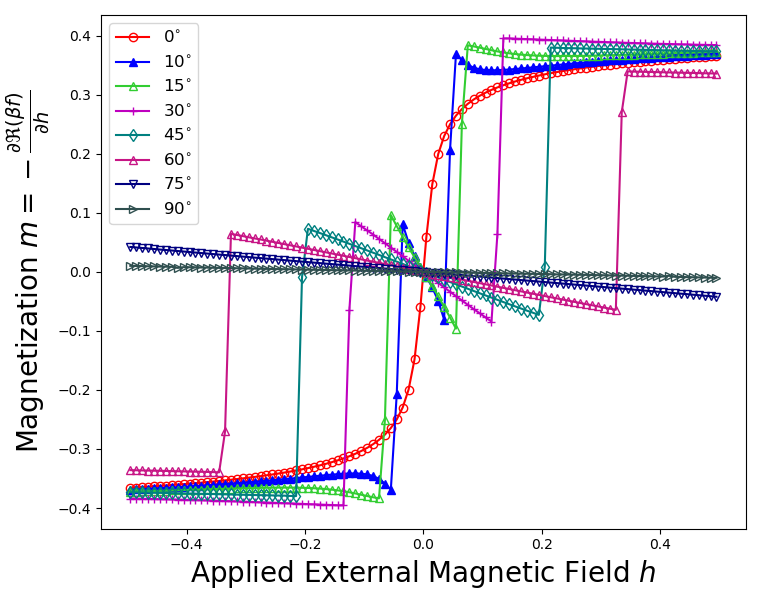}
\end{center}
\caption{Magnetization as a function of applied uniform magnetic field in the two phases of the isotropic 2D Ising model in the complex $\sinh(2 \beta)$ plane.  Upper panel:\ ferromagnetic phase, $\vert \sinh(2 \beta) \vert = 1.3$. Lower panel:\ metamagnetic response in the paramagnetic phase, $\vert \sinh (2 \beta) \vert=0.9$.}
\label{f:magnetization}
\end{figure}

To investigate further the nature of the NFM phases, we perform TRG calculations of the properties of the 2D Ising model in a uniform external magnetic field.  We begin with the isotropic case, where no modulated order is expected.  Here it is convenient to use a slightly different complex-temperature variable, $\sinh (2\beta)$; in the complex $\sinh (2\beta)$-plane, the PM-FM transition of the isotropic model lies on the unit circle.

Our results are presented in Fig.~\ref{f:magnetization}.  The upper panel shows the uniform magnetization as a function of applied field as we follow a contour inside the FM phase, $\vert \sinh (2\beta) \vert = 1.3$, from the real to the imaginary axis.  There is clearly a non-zero spontaneous magnetization at every point along this contour, the magnitude of which depends only weakly on the angle.  The lower panel shows the same kind of plot, but for a contour inside the PM phase, $\vert \sinh (2\beta) \vert = 0.9$.  Here, as expected, there is no spontaneous magnetization\footnote{Note that the low-field regime is apparently diamagnetic. This appears to be a non-universal peculiarity; in some cases we observe similar metamagnetic jumps on top of a more conventional paramagnetic response, including in analytically solvable cases where the transition from paramagnetic to diamagnetic response is clearly seen in the normal phase and is unaccompanied by anything of consequence.}, though there are metamagnetic transitions at finite values of the applied field, as expected from the fact that the usual Ising critical point turns into a first-order transition\cite{Fisher1965} at complex coupling.  We have also tracked the onset of finite spontaneous magnetization at the phase transition point and found \emph{two} distinct types of critical point:\ the familiar real-temperature critical point, but also a `unitary' critical point at $\tanh \beta=\sinh 2\beta = i$, discussed further below.

\begin{figure}
\begin{center}
\includegraphics[width=0.85\columnwidth]{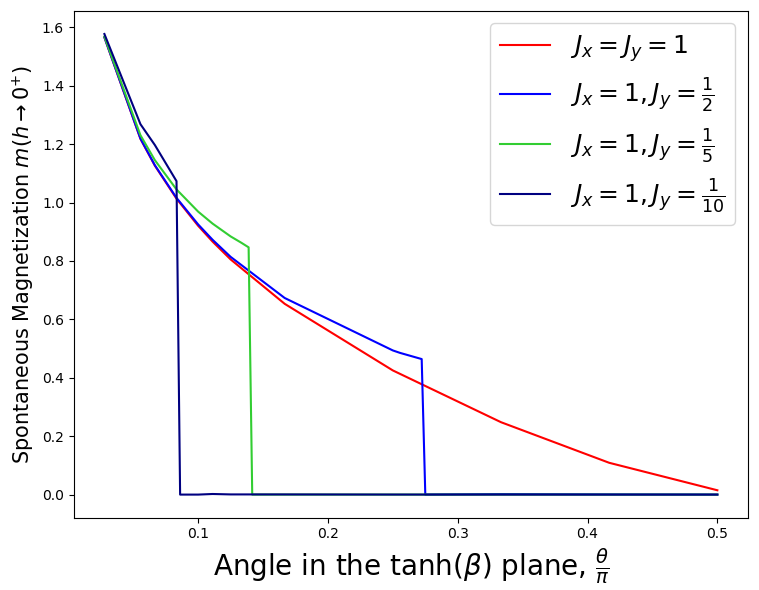}
\end{center}
\caption{The spontaneous magnetization of the anisotropic 2D Ising model as a function of angle from the real-temperature axis on the contour $\vert {\tanh \beta} \vert = 1$, for four different values of the anisotropy $J_y/J_x$.  The jump corresponds to the FM-NFM1 transition, shown for the $J_y/J_x = 1/10$ case in Fig.~\ref{f:phasediag}.}
\label{f:magvsanis}
\end{figure}

Now we turn to the anisotropic 2D Ising model, extracting the spontaneous magnetization using the kind of analysis shown in Fig.~\ref{f:magnetization}.  Fig.~\ref{f:magvsanis} shows our results:\ a plot of the spontaneous magnetization as a function of angle from the positive real axis along the contour $\vert \tanh \beta \vert = 1$, for various values of the anisotropy $J_y/J_x$.  In the isotropic case, there is spontaneous magnetization all the way along this contour, in keeping with the results of Fig.~\ref{f:magnetization}.  In the anisotropic cases, however, the spontaneous magnetization drops abruptly to zero at a critical angle $\theta_c$, which moves closer to the real-temperature axis as the anisotropy is increased\footnote{Oddly, the anisotropy appears to strengthen the ferromagnetic order itself.}.

In summary, the NFM phases in the anisotropic 2D Ising model at complex temperature are separated by critical surfaces from both the paramagnetic phase and the ferromagnetic one, as shown in Figs.~\ref{f:energyspecheat}) and \ref{f:magvsanis}) respectively.  Furthermore, they do not possess any uniform magnetization, as Fig.~\ref{f:magvsanis}) shows.  We conclude that they are long-range ordered, but that their spin-spin correlations are modulated.  This is implied by the above transfer-matrix results on $M$-leg ladders; the claim is also supported by preliminary results on two-point correlation functions\cite{basuprep}.

{\it Unitary points.}
Finally, let us briefly discuss the physics of the unitary points in the complex-temperature plane, $\tanh \beta = \pm i$.  An early motivation for this study was the exploration of dynamics that is quantum in nature yet lacks full unitarity, similar to that of open quantum systems.  In $M$-leg ladders it seems natural to think of the long direction as the `clock' direction, i.e.\ the time direction in the familiar continuum path-integral formulation.  The modulated magnetization observed in the $M=1$ problem is then a simple example of coherent spin precession when $\vert \tanh \beta \vert=1$, with frequency $\arg(\tanh\beta)$).  For $\vert \tanh \beta \vert \ne 1$, this acquires some decay, controlled by the distance from the unit circle in the complex-temperature plane.  For $M>1$ the situation is less coherent, with most of the $2^M$ possible correlations decaying, except for those corresponding to a few dominant eigenvalues.

Remarkably, the complex temperature plane possesses points where the full unitarity of the (properly normalized) transfer matrix is restored, with all $2^M$ eigenvalues becoming unimodular:\ a sufficient condition for unitarity provided that the transfer matrices are symmetric, which ours are by construction.  At these points, all of the correlation lengths are infinite, i.e.\ there is no decay.  This may be checked explicitly for $M$-leg ladders with arbitrary anisotropy, and also holds locally in tensor-network representations of extended 2D problems\footnote{N.\ Pomata {\it et al.\/}, unpublished.}.  As noted above, these unitary points also appear multicritical, with spontaneous uniform and staggered magnetization vanishing as they are approached.

{\it Acknowledgments.}
We are grateful to Sasha Abanov, F. Beichert, W. Bialek, R. Blythe, Michael Fisher, E. Fradkin, S. Gopalakrishnan, D. Huse, S. Kivelson, R. Moessner, G. Mussardo, F. Pollman, N. Pomata, M. Stoudenmire, and Tzu-Chieh Wei for stimulating discussions over the years. CAH gratefully acknowledges financial support from UKRI under grant number EP/R031924/1; he is also grateful to Rice University for a visiting appointment in spring 2019, where part of this work was completed. SB and VO acknowledge support from the NSF
DMR Grant No.\ 1508538 and US-Israel BSF Grant No.\ 2014265.

\end{document}